\documentclass[manuscript]{acmart}

\usepackage{listings}

\AtBeginDocument{%
  \providecommand\BibTeX{{%
    \normalfont B\kern-0.5em{\scshape i\kern-0.25em b}\kern-0.8em\TeX}}}

\copyrightyear{2022} 
\acmYear{2022} 
\setcopyright{acmcopyright}\acmConference[PEARC '22]{Practice and Experience in Advanced Research Computing}{July 10--14, 2022}{BOSTON, MA}
\acmBooktitle{Practice and Experience in Advanced Research Computing (PEARC '22), July 10--14, 2022, BOSTON, MA}
\acmPrice{15.00}
\acmDOI{10.1145/3311790.3400848}
\acmISBN{978-1-4503-6689-2/20/07}

\begin{document}

\title{A Framework to capture and reproduce the Absolute State of Jupyter Notebooks}

\author{Dimuthu Wannipurage}
\email{dwannipu@iu.edu}
\affiliation{%
  \institution{Cyberinfrastructure Integration Research Center,}
 \institution{Indiana University}
  \city{Bloomington} 
  \state{IN}
  \country{USA}
   \postcode{47408}}

\author{Suresh Marru}
\email{smarru@iu.edu}
\affiliation{%
  \institution{Cyberinfrastructure Integration Research Center,}
 \institution{Indiana University}
  \city{Bloomington} 
  \state{IN}
    \country{USA}
   \postcode{47408}}
   
 \author{Marlon Pierce}
\email{marpierc@iu.edu}
\affiliation{%
  \institution{Cyberinfrastructure Integration Research Center,}
 \institution{Indiana University}
  \city{Bloomington} 
  \state{IN}
    \country{USA}
   \postcode{47408}}

\renewcommand{\shortauthors}{Wannipurage et al.}
\renewcommand{\shorttitle}{Recreatable Jupyter Kernel}

\begin{abstract}
Jupyter Notebooks are an enormously popular tool for creating and narrating computational research projects.  They also have enormous potential for creating reproducible scientific research artifacts. Capturing the complete state of a notebook has additional benefits; for instance, the notebook execution may be split between local and remote resources, where the latter may have more powerful processing capabilities or store large or access-limited data.  There are several challenges for making notebooks fully reproducible when examined in detail.  The notebook code must be replicated entirely, and the underlying Python runtime environments must be identical.  More subtle problems arise in replicating referenced data, external library dependencies, and runtime variable states.  This paper presents solutions to these problems using Juptyer’s standard extension mechanisms to create an archivable system state for a running notebook. We show that the overhead for these additional mechanisms, which involve interacting with the underlying Linux kernel, does not introduce substantial execution time overheads, demonstrating the approach's feasibility. 
\end{abstract}



\maketitle

\section{Introduction}
Project Jupyter’s notebook format \cite{kluyver2016jupyter} has revolutionized interactive computing and has become ubiquitous among researchers using machine learning techniques and in scientific computing communities. Jupyter’s simple-to-use user interfaces, ease of deployment, rich visualization support, and support for multiple programming languages have nurtured a large user community with a diverse set of use cases, from simple python code execution to complicated neural network simulations in high-performance computing environments. When code execution and visualization are combined with embedded textual descriptions, the Jupyter ecosystem for computing and data analysis can be viewed as an infrastructure for providing narration for the computing and data life cycle as stories \cite{perez2015project,granger2021jupyter}. This storytelling is essential for reproducible science \cite{national2019reproducibility} or, even better or re-creatable, or tweakable science \cite{rescience}.

The high user adoption of Jupyter has resulted in a large ecosystem supporting multiple programming languages through kernel extensions and integration into users' development environments. Commercial cloud platform integrations such as Google Colaboratory \cite{GoogleColab} and a GitHub-based notebook repository MyBinder have fostered user sharing and collaboration of notebooks. Community support for popular data platforms has also enabled notebook reuse \cite{JupyterKernels}.

Reproducibility is an essential aspect in any science discipline and sharing someone’s work through a notebook may be only meaningful if others can reproduce their work. While there has been a significant amount of work done in making Jupyter Notebooks configurable and accessible to resonate with the user’s requirements, there has been significantly less work to make notebooks reproducible across various platforms. We have identified several areas where reproducing a notebook execution is critical based on these considerations.

\section{Use Cases for Portable and Reproducible Notebooks}

\subsection{Reproducing Code from Scientific Publications}
Software is a critical component of science, and reproducibility \cite{national2019reproducibility, jay2020software, peng2011reproducible}, and scholarly articles are increasingly referencing Jupyter Notebooks with code segments and visualizations. However, it is very cumbersome if not impossible to reproduce the published bundle of notebooks and associated code and data. Few challenges include recreating implicit data and identical runtime environments, including a cascade of runtime library dependencies. Notebook Kernels such as IPython rely on local python virtual environments. Notebooks also interact with accessible external computational and data resources that are not easily translatable to shared collaborators. Out of band communication with authors of notebooks is necessary to fully reproduce the combination of data, dependencies, and the runtime environment of the published work. Conducting these steps is a non-trivial and time-consuming process.

\subsection{Sharing Classroom Learning Material }
Jupyter Notebooks are a popular mechanism for providing classroom learning materials. Examples such as \cite{chastang2018unidata} allow Instructors to share notebooks, and students can execute them on their laptops during a  classroom session. There are several cases when this process becomes significantly complicated. Students’ laptops will have various operating system types and versions. Some may have different Python runtime versions, which may be incompatible with the instructor's notebook version. In some cases, instructors might need to use external data files as inputs for the code fragments in the notebook, and these also need to be shipped with the notebook code. 

\subsection{Transferring Local Computations to Between Local and High-Performance Computing Environments}
Researchers have access to a scale of computing power from laptops to high-end supercomputers and computational clouds.  With the profound usage of Jupyter, it is essential to support notebook portability across these resources without fragmenting data and code too much. As an illustrative use case, a researcher might start coding a notebook on a workstation but training a higher-order neural network may necessitate a computer with a powerful GPU, large memory, and access to large data sets. Potentially post-processing analysis may be bought back to a local laptop. In other cases, portions of the notebook may need to be run on a computer containing the data that cannot be easily transferred to another place due to higher data volume or low network bandwidth. The data may need to remain under specific security controls in some cases. 

These mix and match executions within a single Jupyter Notebook are not supported in the standard distribution. Some work has been done in this area \cite{contextawaremigration}, to customize the vanilla iPython kernel and export the code execution to remotely hosted iPython kernel environments. However, these approaches require a significant amount of effort and expertise in distributed systems from the user. In addition, these approaches focus on keeping the Python runtime context synced between external kernel environments but do not address any data level dependency with the code fragments in the notebook. For example, when developing a neural network in a notebook, we use the training and testing datasets available in the local disk. Suppose others need to replicate the same environment, relevant input files must be manually selected. Then the notebook code must be updated with relative file paths and distributed with the notebook. 

\subsection{Running on On-Demand or Time-Bound Computing Infrastructure}\
Preemtable infrastructures like Amazon Spot Instances \cite{spotinstances} or HPC shared resources with bounded maximum wall time do not guarantee the complete end-to-end execution of a long-running notebook. Users may also need to stop the notebook execution and create a snapshot to resume it from the last checkpoint at another time by reloading all the data and local context of the Jupyter session. In such scenarios, having a feature to create a comprehensive snapshot of the runtime saves time and computing resource utilization. When Colab Notebooks \cite{GoogleColab} expire after the threshold, the platform kills the notebook, and all the local states are lost unless the user manually serializes and uploads it to Google Drive. It would be beneficial to run the notebook from the last stopped checkpoint even after the original session has expired in all of these cases.

\section{Reproducibility of Notebooks}

Jupyter Notebooks inherently are sharable, but several missing components hinder the process of reproducing results in a different environment. Restarting a running notebook in a different environment can be made possible by five essential requirements: 

\begin{enumerate}
    \item The code within the notebook is portable. 
    \item Python runtimes should be identical. 
    \item Replicate referenced local data.
    \item External library dependencies should be replicated or installed if missing. 
    \item Python runtime variables should be re-initialized.
\end{enumerate}

\begin{figure}
\centering
\includegraphics[width=300px]{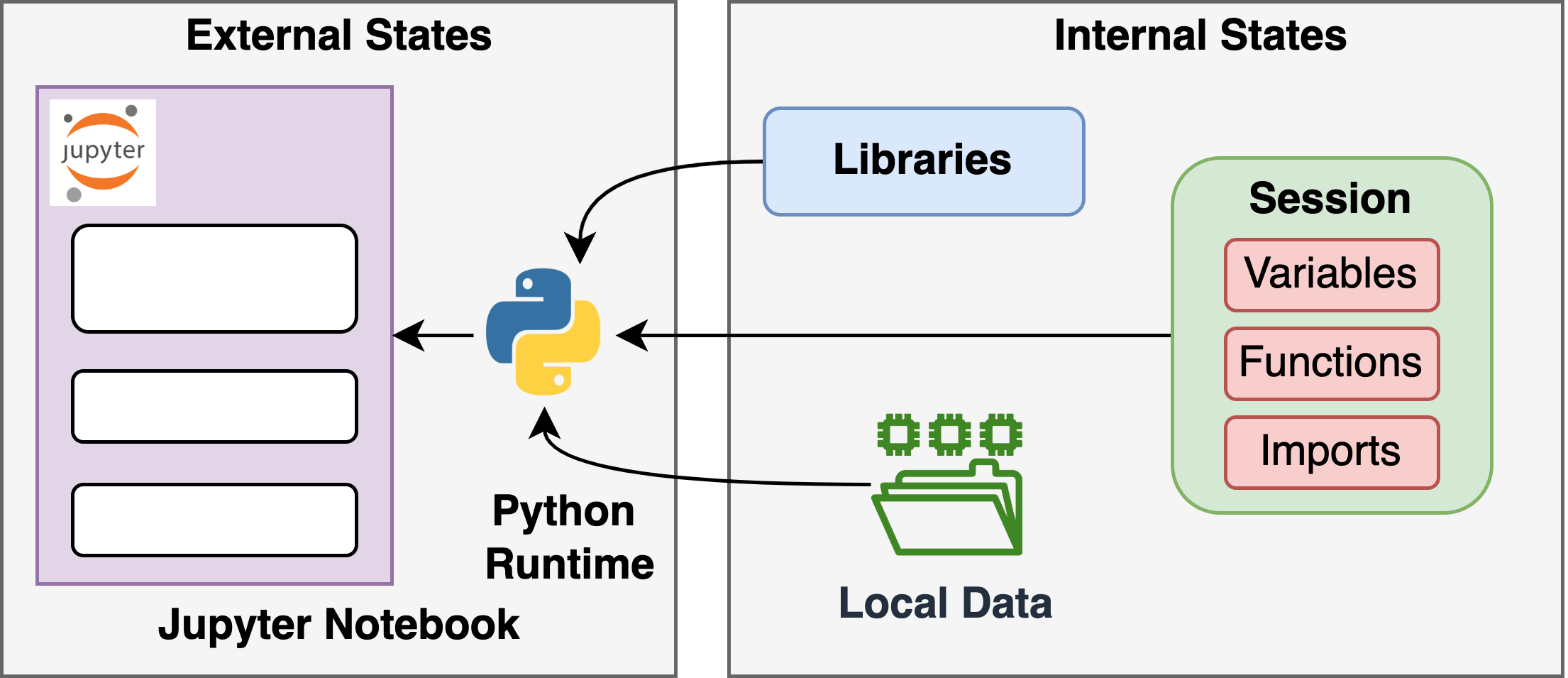}
\caption{Components associated with the full state of a Jupyter Notebook. External states are not associated with the execution, and internal states depend on the code and execution order of the notebook.}
\label{fig:fig1}
\end{figure}

Requirements 1 and 2 are external states which do not depend on the execution of the notebook. Requirements 3, 4, and 5 are internal to the Python runtime and the order of cell execution of the notebook (Figure 1). If any of the above five requirements are not met, the replicated environment is not identical to the source environment. In most sharing solutions \cite{tensimplerules, NeurophysiologicalAnalytics} we can only see that requirements 1 and 2 are satisfied but not requirements 3, 4, and 5.

In this paper, we propose and implement a framework for Jupyter Notebooks to satisfy all five requirements when replicating the state of a running notebook. In addition to the functional requirements, our solution is designed to prioritize the ease of use for users so that they do not have to put any more effort into making this work than running a vanilla notebook environment. 

\section{A Novel Approach to Capture the Absolute State of Notebooks}

To achieve the absolute state capture of a notebook, we primarily utilized the Jupyter Magic Extension support provided by the IPython kernel. In addition, we made modifications to the iPython kernel to integrate with external services to capture information that cannot be directly captured from the Kernel itself. We did not focus on steps 1 and 2 in the framework development as they are tightly coupled with the deployment. Current framework implementation, which is relevant to the scope of this paper, is available in the GitHub repository \cite{githubcode}. 

\subsection{Capturing Data Dependencies}
Suppose a Python code in a Jupyter Notebook uses a file in the local file system for reading or writing purposes. In this case, this event should be captured if we want to replicate the data dependencies. However, this is a highly complicated task. There may be many Jupyter session-to-file system interactions either directly through the code the user has written or implicit libraries and function calls of the Python runtime. For example, a simple “pip install” command might access hundreds of files in the pip cache, and we should not consider them as data dependencies for the replication. Assuming that we have a way to filter out unwanted file system events, we still have the issue of capturing these file system events. Python and the iPython kernel do not have any native API to monitor file operations, so we must look for alternate ways.

\subsubsection{Monitoring Linux Kernel System Calls by the  IPython Kernel: }
All file operations are forwarded to the operating system as system calls regardless of the language-level file system operation implementation. To be precise, in Linux, all of these language bindings invoke the ``openat’’ \cite{openatsyscall} system call to get the file descriptor of the file. We decided to capture the events from the Jupyter kernel process to the Linux kernel and log the invocations targeting ``openat’’ system calls. We used Linux’s ``strace’’ \cite{stracecommand} tool to filter and log the ``openat’’ system calls coming through the IPython kernel process, which is responsible for the target Jupyter Notebook. There is a one-to-one relationship with a running Jupyter Notebook session and an IPython kernel process by design, so it is easy to derive file operations done by a notebook by capturing the procedures done by the designated IPython kernel process. Through this method, we managed to capture all the file system events coming from the Jupyter kernel runtime process irrespective of the language bindings used to access data at the programming language level.

The integration of the IPython kernel tracing framework is performed in two separate steps. In the first step, we capture system calls sent by each IPython kernel and log them into individual process-specific log files. In the second step, we analyze these logs files and identify relevant file system invocations.

Fig.2 illustrates the first stage processes. We developed a process-tracing server, which can start Linux ``strace’’ subprocesses for monitoring given process ids. These ``strace’’ sub-processes capture the ``openat’’ system calls and log them in a specified log file. When a user creates a Jupyter Notebook, the Jupyter Notebook Server initializes a dedicated IPython kernel process to handle requests. We updated the start method of the IPython kernel to invoke the process tracing server with its process id as a parameter. The communication between these components is performed through a Unix socket to minimize the communication overhead. Once the process tracing server receives the kernel start message, it starts a ``strace’’ process to trace the IPython kernel process. Dashed arrows in Fig. 2 show the tracing connection created from the ``strace’’ job. Once the events are captured, the ``strace’’ job writes only ``openat’’ system call log to a designated log file. Dotted arrows in Fig.2 depict that connection. ``strace’’ Linux command needs to be executed as the superuser to access kernel system calls. For this reason, we separate the IPython kernel and process tracing servers into two separated entities; IPython kernels run as the regular user and the tracing service runs as the superuser.

\begin{figure}
\centering
\includegraphics[width=350px]{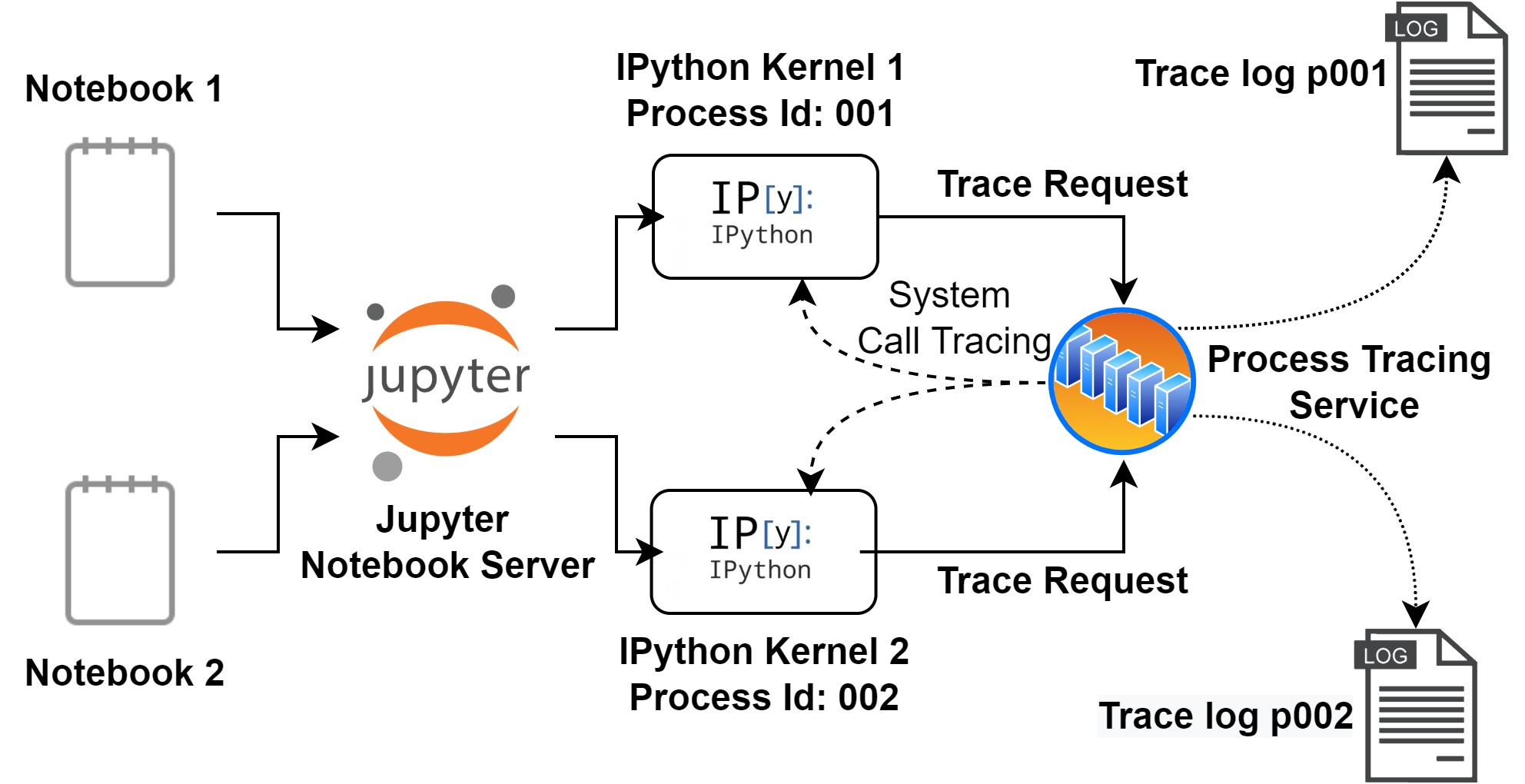}
\caption{The workflow for capturing the filesystem calls from a Jupyter Notebook process to the operating system kernel.  When an IPython process is started for a notebook, the process tracing service receives a notification to trace the process and then logs output into a file. Each notebook process has a different trace log file.}
\label{fig:fig2}
\end{figure}

The second major step analyzes these log files and captures relevant file system operations to create a list of files that the Jupyter Notebook accessed. To achieve this, we developed a Jupyter Magic \cite{jupytermagiccustom} extension to parse the traced log file and filter out relevant file operations for further processing (Fig. 3). At any point in the notebook, users can run the Jupyter Magic line command to invoke this workflow and get a list of files that they directly or indirectly used in the notebook up to that point. We use a rule-based filter to ignore all operations not required to reproduce the notebook state. Discounted file operations include those associated with Python cache files, Jupyter Notebook checkpoint files, library files created and accessed as byproducts of notebook autosaves, and pip install commands for filtering relevant file operations.

\begin{figure}
\centering
\includegraphics[width=350px]{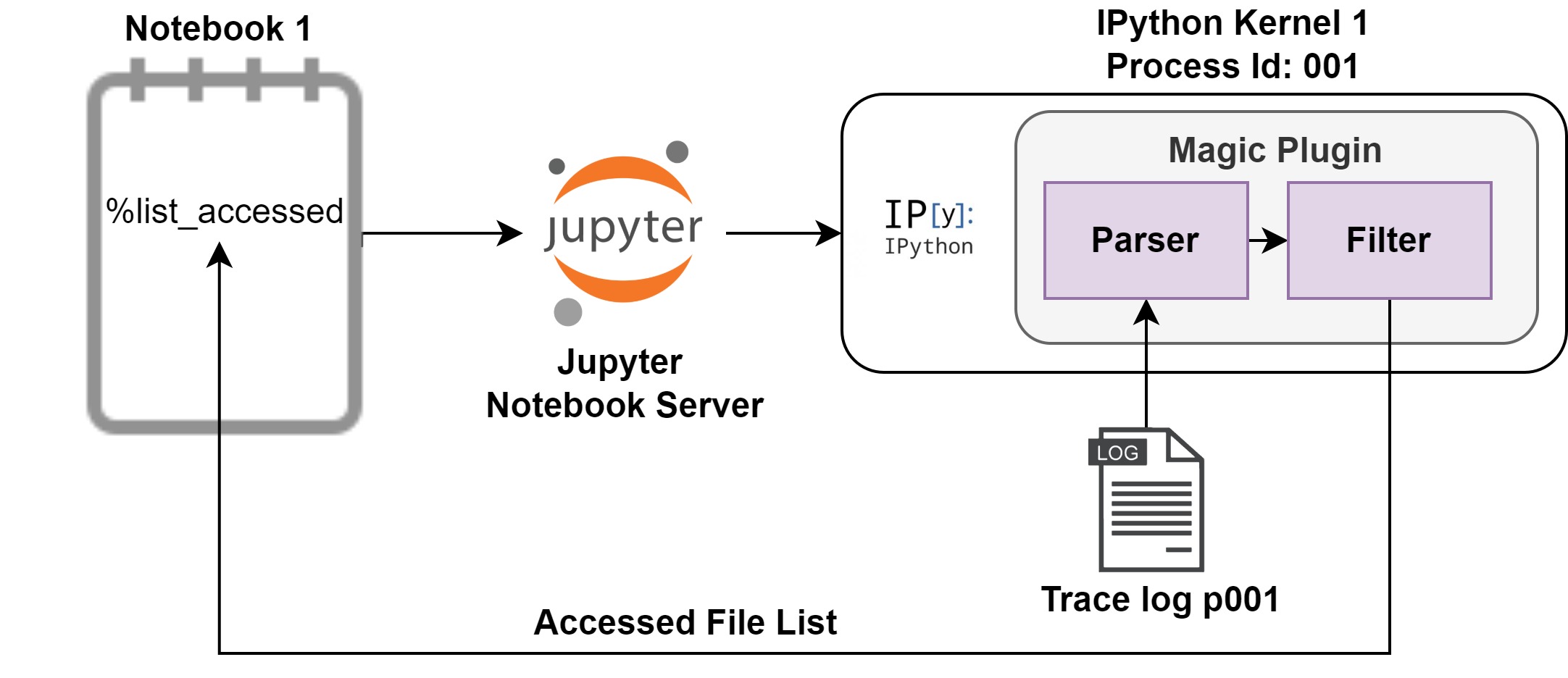}
\caption{Workflow of processing a trace log file inside a Jupyter Magic plugin to shortlist the valid list of files accessed by the Jupyter Notebook. Trace log p001 is the log file associated with Notebook 1 that is created as depicted in Fig. 2.}
\label{fig:fig3}
\end{figure}

These filtered files are being used as the reference to create the export of data dependencies for the notebook execution and it is explained in the next section of the paper.

\subsection{Determining  Dependencies on Optional Libraries}
Python’s runtime ships with a default set of libraries like math, io, and sys to perform common operations for mathematical, system, and file operations. In addition, many out-of-the-box libraries can be installed from the pipi.org repository \cite{pypi}. Typically users use these libraries in notebooks either by installing them inside the notebook using the “pip install” command or using already installed libraries from the Python library cache of the running machine. However, each library may have multiple versions that contain some API level and performance-related changes, so the code written using a particular library is somewhat coupled with the version of the library being used. When we share a notebook containing these library imports, there is no specific way to guide the users on which libraries and versions need to be installed before running them. Typically, the approach is to run the shared notebook, and if there is a missing library in the Python environment, install the missing one using the “pip install” command. However, it is still not guaranteed to work properly as we do not know the exact version of the library that was installed in the author’s environment. For this reason, syntax errors and performance issues are possible even though the relevant library is installed in the second system.

Considering the above concerns, deriving the optionally installed libraries and their versions loaded in a Jupyter notebook session is critical to reproducing the library dependencies. However, this needs to be carried out carefully. In Python, the library version loaded into the notebook’s IPython session depends on the way it is being bootstrapped. If the notebook was installed in a Python virtual environment, dependencies are loaded from the virtual environment’s context. If, instead, the notebook is executed in the user’s default environment, libraries are loaded from Python's global library cache. The best way to derive the correct library version is to connect to the specific notebook session and get the relevant library version. 

To solve this problem, we developed another Python Magic extension that can be bundled into the running IPython kernel of the notebook. According to the IPython Magic framework, all the extensions receive the local namespace of the running IPython process. We can find all the imported libraries and versions inside this local namespace using a type-based filter below. 

\begin{lstlisting}
def imports():
   for name, val in local_ns.items():
       if isinstance(val, types.ModuleType):
           yield val.__name__
\end{lstlisting}

Next, we filtered out default libraries shipped with the Python runtime and captured the externally installed libraries. Once we get the library names, versions of the loaded library can be derived using the importlib \cite{importlib} library.

\begin{figure}
\centering
\includegraphics[width=350px]{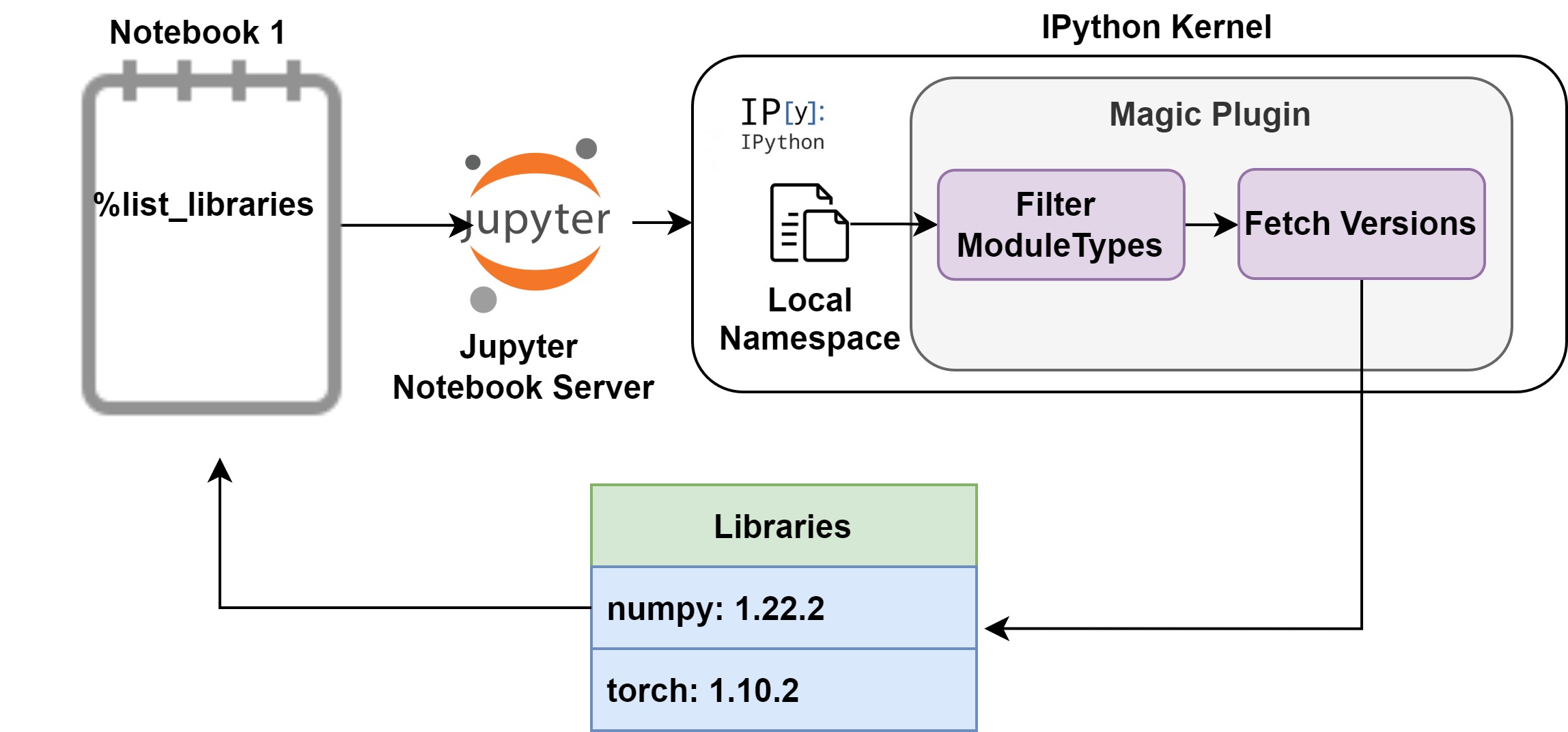}
\caption{Magic plugin extensions capture the libraries' list and their versions imported inside a notebook session.}
\label{fig:fig4}
\end{figure}

Once all the relevant library names and versions are fetched, we save that metadata in a Python dictionary object where keys are the library name and values are versions (Fig. 4). This dictionary is used to export the library dependencies, as described in Section 5.

\subsection{Capturing the Notebook’s Runtime Session}
The last and most challenging step to achieve complete reproducibility of a Jupyter Notebook is to capture the notebook’s runtime session. This is an essential feature for checkpointing a notebook and restarting it in the same or different environment to continue from the last executed position. This requires serializing the execution session and reloading it back when it is restarted. A typical running Jupyter session may include library imports, initialized variables, and function declarations. If we want to create a snapshot of a session, we need a way to serialize all these types of entities when storing the state and deserialize them when restoring to a running session. In addition, we need to find a way to programmatically list the minimal set of entities that is sufficient to replicate the entire session.

A local namespace is a dictionary of objects where keys are entity names and values are entities created throughout the notebook session. These keys consist of variable names, function names, imports, and previous cells’ execution ids. We developed a Jupyter Magic extension (Fig. 5) to capture this session information using the local namespace information that the extension receives from the Jupyter framework. Here we are only interested in the global variables, function definitions, and library imports required to recreate a session. 

The challenge is to derive the subset of these entities from the local namespace as it represents all the entities as generic key-value pairs. To find this set of entity names, we use the ``who’’   \cite{whomagic} built-in Jupyter Magic command, which prints the signatures of imports, functions, and global variables. Once those entities are identified, we create a minimal version of the local namespace dictionary by filtering derived entity names. To export this sub-namespace object, we serialize it to a binary file using the ``dill’’ \cite{dilllib}  library, which is capable of serializing most of Python objects into binary format. However, there may be scenarios where some objects in the sub-namespace can not be serialized. To address this problem, we do a binary search on the sub-namespace and perform serialization tests on each subtree to determine the availability of any possibly incompatible entity. We remove any incompatible entities from the sub-namespace object and warn the user if there are any.

\begin{figure}
\centering
\includegraphics[width=350px]{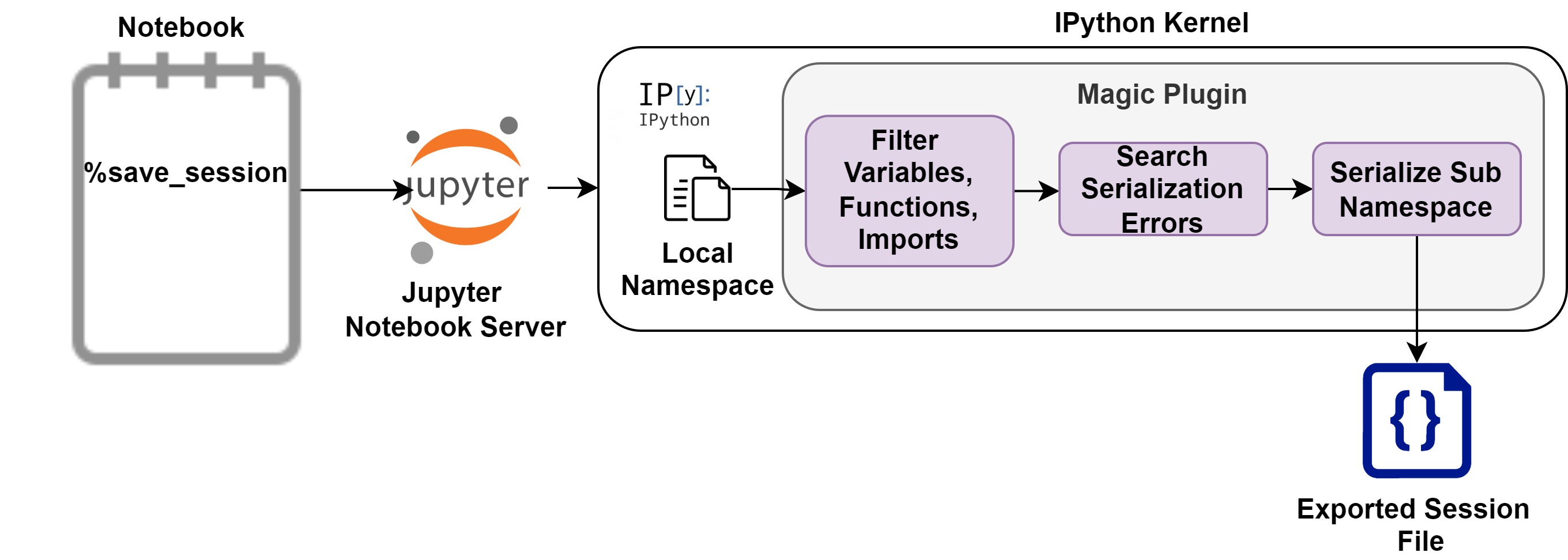}
\caption{Magic Plugin implementation workflow to capture the variables, imports and functions in the local namespace of the Jupyter Notebook session and serialize into a file.}
\label{fig:fig5}
\end{figure}

Once the sub-namespace object is serialized into byte format, we use it to build the notebook state bundle as mentioned in Section 5

\section{Strategy for Reproducing Stateful Notebooks}
The previous section discussed how we capture the information to reproduce the whole notebook state. This section outlines how we package and export that information to be reproduced in a different environment. We also discuss how to bootstrap an environment when we import a notebook state from an exported bundle as mentioned above.

\subsection{Packaging Strategy for Captured States}
We use an archiving approach when exporting a notebook state that uses all of the above-mentioned state capturing techniques. First, all the files identified in the data dependency capturing stage are copied into the root level of the archive directory. To avoid conflicts, each file is duplicated with a unique UUID name rather than the original name. We also create a JSON file containing UUID-to-absolute-file-path mappings; this JSON file is also copied to the root level of the archive directory and the data. Library dependencies that we capture (as described in the previous section) are converted into a JSON format that contains library names and versions; this JSON file is also placed in the root level of the archive directory. In the notebook session capturing stage, we serialize the sub-namespace dictionary object to a file; it is also copied into the root level of the archive directory. Finally, we copied the notebook file as the last stage of archive creation. Figure 7 illustrates the contents and file structure of the archive. 

\begin{figure}
\centering
\includegraphics[width=350px]{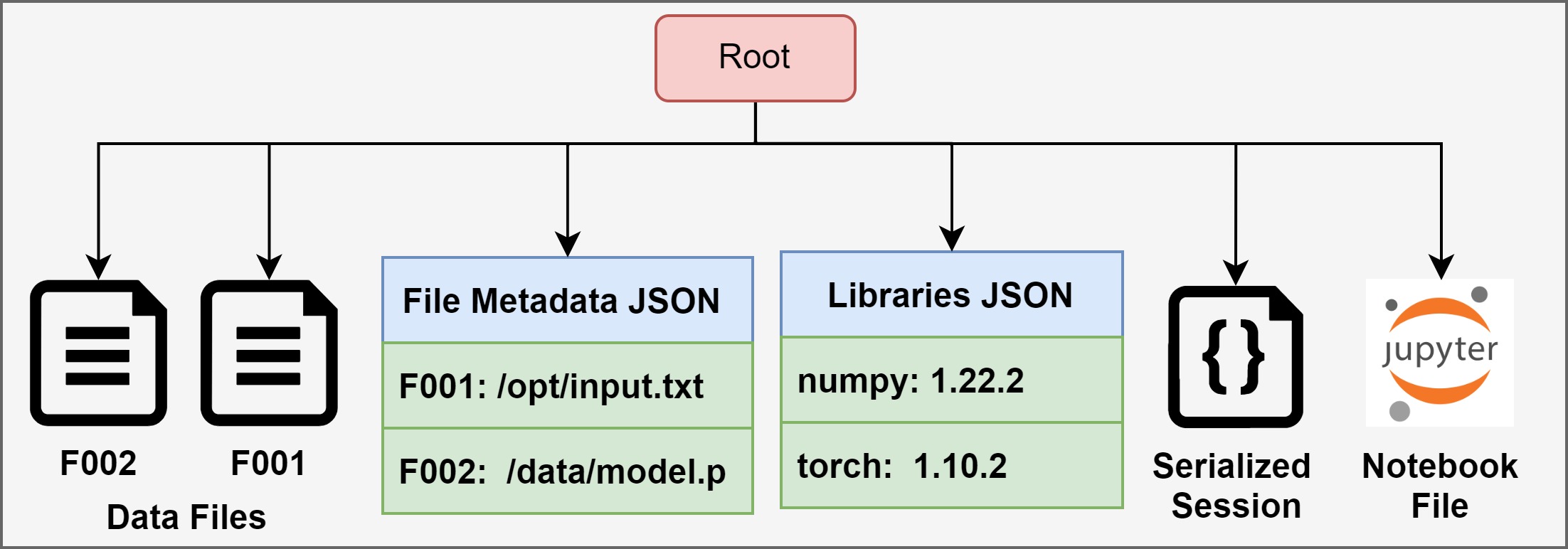}
\caption{ The structure of the archive file that contains a serialized stateful Jupyter Notebook }
\label{fig:fig6}
\end{figure}

To simplify the process, all of the operations mentioned above are integrated into a single Jupyter Magic command so that the user can invoke it inside the notebook file. Upon invocation, the custom Magic code internally invokes the state capturing magic commands described earlier, creates the archiving directory based on the data retrieved from those invocations, creates the archive file using the zip tool, and provides a link for the user to download it through the browser. This archived zip file can be later used to recreate the same state of the notebook in a different environment.

\subsection{Recreating a Stateful Notebook from an Exported Archive}
To recreate a notebook environment from an exported archive, we need to execute the following steps.

\begin{enumerate}
    \item Unarchive the exported archive file.
    \item Parse the File Metadata JSON file and copy data files into designated locations provided in the values of the JSON.
    \item Parse the Libraries JSON file and install those dependencies with versions using pip install commands.
    \item Start the IPython kernel and initialize a session to the exported notebook files.
    \item Run a Jupyter magic command to parse the serialized session and merge it with the local namespace of the notebook.
\end{enumerate}

All of these steps can be performed in any notebook environment, including bare metal, cloud, and containerized platforms. We have developed a single script to accept the archive file with the notebook runtime to reproduce the archived environment. However, there may be issues when placing data files in Step 2 if the notebook is running in a different operating system that does not follow the file system structure of the source environment. For example, a notebook archive created on a Linux computer may fail at Step 2 on a computer running a Windows operating system.  

Considering these scenarios, we recommend using a containerized environment when reproducing a notebook to make the initialization process independent of the host’s file system. Both state-recording and state-replicating notebook environments can be bootstrapped from a one-line command to make the process as simple as possible for the user. To facilitate state recording and recreate a notebook, we have developed a container image that bundles a Jupyter runtime \cite{customdockerimage}, a tracing server, and a python installation. 

To start a notebook environment in state recording mode, users can run the command:

\begin{lstlisting}
docker run --cap-add=SYS_PTRACE -it -p 8888:8888 dimuthuupe/ipykernel:1.0
\end{lstlisting}

To recreate an environment from a state archive:
\begin{lstlisting}
docker run --cap-add=SYS_PTRACE -it -p 8888:8888 
-v <ARCHIVE_FILE>:/opt/ARCHIVE.zip dimuthuupe/ipykernel:1.0
\end{lstlisting}

\section{Performance analysis}

We use an external Linux ``strace’’ process to capture file operations performed by the notebook process to export the file list required to reproduce an identical environment. However, running an ``strace’’ process along with the notebook to capture system calls may have performance implications because recording the operating system’s system calls will cause them to block or wait. We only record ``openat’’ system calls at the design level to mitigate the impact. However, there may still be a performance hit as everything from a mounted disk to a network socket in Linux is represented as files. We may be capturing unwanted events that will eventually affect the notebook’s overall performance.

We performed various stress tests on strace-enabled and disabled environments to evaluate this impact. We ran IO intensive, CPU intensive, and mixed workloads to identify areas affected by a bottleneck. All the performance tests were carried out on a workstation with 32 GB memory, 16 cores and disk write speed of 1GB/s. To keep both environments consistent, we ran both notebooks as Docker containers.

To enable and disable stracing, the container distribution of the kernel accepts the environment variable \verb|ENABLE_TRACE| as a runtime configuration.


\subsection{Running IO-intensive workloads}
\subsubsection{Single file write performance: }
In this experiment, we carried out a test to measure the performance impact from tracing system calls for a single file write. To simulate this scenario, we ran the following Python code fragment with various file sizes to measure the time it took to write a file into the disk with random characters. We captured CPU time and overall wall time for each run as metrics. CPU time is the time the code spends in the CPU, including the Linux kernel, and wall time is the total time it takes to run, including the CPU time and system call overheads (Fig. 7).

\begin{lstlisting}
with open("random.dat", "wb") as output:
    output.write(np.random.bytes(FILE_SIZE))
\end{lstlisting}

\begin{figure}
\centering
\includegraphics[width=\linewidth]{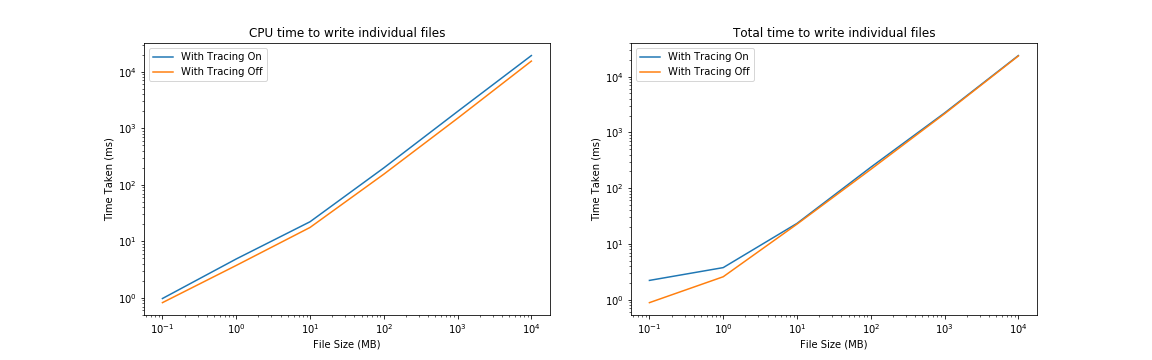}
\caption{ CPU Time and Total Time to write a single file  in various sizes using the Notebook’s Python runtime in tracing on and off scenarios }
\label{fig:fig7}
\end{figure}

\subsubsection{Multiple file write performance: }
Then we ran the same logic of writing files parallelly across 16 cores to measure the performance of parallel file writes (Fig. 8). This experiment fully utilizes CPU, memory, and disk IO, including multi-core executions, in-memory buffer creation, and parallel disk writes. 

       	 

\begin{figure}
\centering
\includegraphics[width=\linewidth]{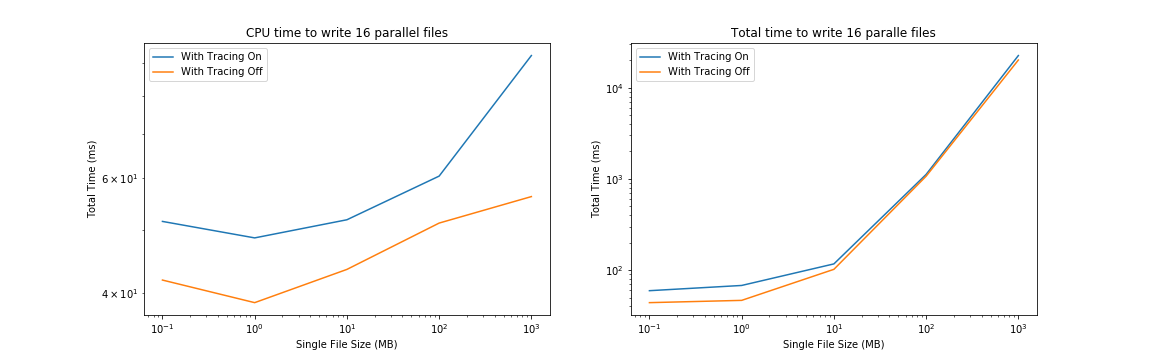}
\caption{  CPU Time and Total Time to write 16 parallel files (1 per CPU Core) in various sizes using the Notebook’s Python runtime in tracing on and off scenarios }
\label{fig:fig8}
\end{figure}

\subsection{Running CPU Intensive workloads}
To measure how the system performed when the CPU is fully utilized, we ran static counters on each CPU core parallelly using the following code fragment and measured the time taken for all the counters to finish (Fig. 9).

\begin{lstlisting}



def f(x):
  counter = 1
  while True:
    counter = counter + 1
    if counter > COUNTER_SIZE:
      break
Pool(psutil.cpu_count()).map(f, range(processes))
\end{lstlisting}

\begin{figure}
\centering
\includegraphics[width=\linewidth]{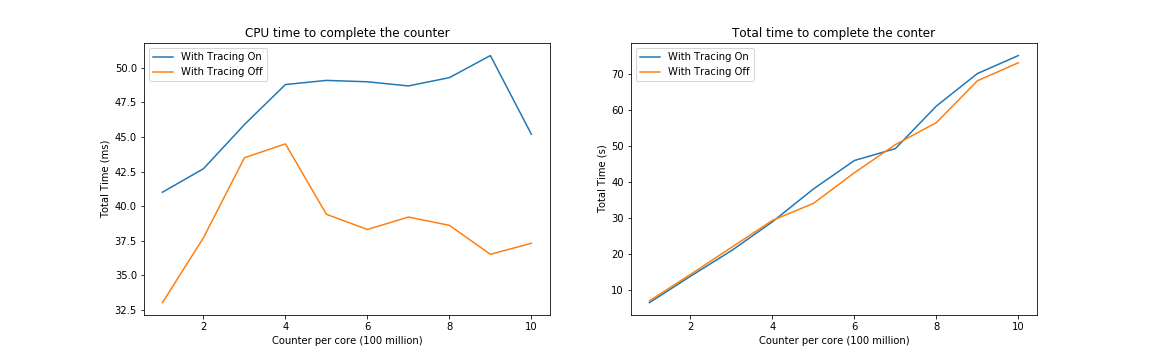}
\caption{  CPU Time and Total Time run parallel counters (1 per CPU Core) in various sizes using the Notebook’s Python runtime in tracing on and off scenarios }
\label{fig:fig9}
\end{figure}

\subsection{Observations}
We have analyzed 3 test scenarios: IO intensive, mixed (both IO and CPU), and CPU intensive operations to measure the impact of tracing the Jupyter notebook kernel process using “strace” program to capture file operations. Among all the 3 scenarios, we observed that IO intensive operations are the least affected by strace, while CPU intensive operations are the most affected in terms of the CPU time spent. However, measuring the total time to complete these operations, which is the real time between start and end, shows that significantly less cost is paid by the tracing enabled setup than a non-tracing environment. It is very clear that tracing affects CPU time, but because the proportion of CPU time to the total time is very low (the majority of total time is governed by the system call overhead), the impact  of the CPU time is not significantly affecting the final overall performance.

\section{Conclusion \& Future Work}
This paper presents  the framework-level concepts needed to implement a fully reproducible Jupyter Notebook environment and the results of a performance analysis of possible bottleneck scenarios. Currently, the proof of concept implementation of the framework supports Docker-based executions in single-user mode. Even with this limitation, our approach is potentially  beneficial for long-running notebooks, especially on HPC resources: Users  can pause running notebooks  at some point and resume in a different environment with zero configuration changes. In the future, we plan to integrate this framework as a secure multi-tenanted service on the Jetstream 2 \cite{hancock2021jetstream2} cloud computing system with the scaling support of JupyterHub \cite{jupyterhub}. Integrating with Apache Airavata components, we plan to facilitate seamless remote execution of machine learning models on Jetstream2’s GPU resources coupled with analysis on local resources all within a single notebook session. 


\bibliographystyle{ACM-Reference-Format}
\bibliography{pearc22-jupyter-airavata-kernel}

\end{document}